# Time-refraction and time-reflection above critical angle

# for total internal reflection

Lior Bar-Hillel[1+], Alex Dikopoltsev[2,3], Amit Kam[2], Yonatan Sharabi[2], Ohad Segal[1],

Eran Lustig[2] and Mordechai Segev[1,2]

1. *Department of Electrical and Computer Engineering, Technion, Haifa 32000, Israel*
2. *Physics Department, Technion, Haifa 32000, Israel*
3. *Institute of Quantum Electronics, ETH Zurich, 8093 Zurich, Switzerland*

**Abstract**

We study the time-reflection and time-refraction of waves caused by a spatial interface with a medium undergoing a sudden temporal change in permittivity. We show that monochromatic waves are transformed into a pulse by the permittivity change, and that time-reflection is enhanced at the vicinity of the critical angle for total internal reflection. In this regime, we find that the evanescent field is transformed into a propagating pulse by the sudden change in permittivity. These effects display enhancement of the time-reflection and high sensitivity near the critical angle, paving the way to experiments on time-reflection and photonic time-crystals at optical frequencies.

Modulating the electromagnetic (EM) properties of a medium at ultrafast time scales [1] is now gaining renewed interest due to recent advances in ultrafast switching in highly nonlinear materials [2–7]. Importantly, inducing an abrupt temporal change in the EM properties of a medium is fundamentally different from an abrupt change in space (an interface) because causality plays a crucial role. In the context of light-matter interactions, strong and abrupt changes in the refractive index result in time-reflection and time-refraction [1,8,9], and can yield a variety of phenomena ranging from fast switching of ultrastrong coupling [10–12] and localization by temporal disorder [13] to enhanced emission by dipoles [14], quantum fluctuations and free electrons [15] in photonic time-crystals (PTCs) and time-varying dielectric media [16–21]. PTCs, photonic structures whose EM properties are varied periodically in time with a period comparable to a single cycle of a wave propagating therein, are perhaps the most promising manifestation of such strong abrupt variations in the refractive index [14,22–28]. As we show below, a wave incident upon a spatial interface with a time-varying medium exhibits unique properties.

When an EM wave propagates in a medium whose refractive index varies within a few-cycles, the wave experiences refractions and reflections known as "time-refractions" and "time-reflections" [9]. When the medium is homogeneous, both time-refraction and time-reflection are manifested in the translation of the temporal spectrum, as a consequence of momentum conservation. The time-refracted wave continues to propagate with the same wave-vector, the time-reflected wave is propagating backwards with a conjugate phase (due to the sign change in the frequency) [8,29]. Importantly, while time-refraction is always significant, for the time-reflection to be measurable, the index change has to be large (order of unity) and abrupt (occurring within 1-2 optical cycles), otherwise the time-reflection is extremely weak, and PTCs become unfeasible. This tough requirement to have a strong and abrupt change in the refractive index is the reason why time-reflection of light has never been observed at optical frequencies. Thus far time-reflection was observed with water waves [29] and cold atoms [30] and was proposed in synthetic dimensions [31], but with EM waves, it was observed only at microwaves frequencies [32–34]. The reason it is extremely hard to observe time-reflections at optical frequencies is profound: many nonlinear optics effects are instantaneous (e.g., the optical Kerr effect), but the index change they yield is at least a thousand-fold too weak to cause measurable time-reflections. At the other extreme, some nonlinear effects can provide huge index changes, but their response is orders of magnitude too slow to drive time-reflections because they require transport (of charges, of atoms, etc.). There are only a handful of exceptions: mechanisms giving rise to large index changes occurring within a few femtoseconds (fsec). One of those involves transparent conductive oxides, where recent work has demonstrated index changes of ~0.3 within 5-8 fsec [35–37]. Those experiments showed that there is indeed a mechanism that makes time-reflections and PTCs feasible at optical frequencies. However, even in that experiment – time-reflection was too weak to be measured. These recent experimental results imply that new approaches are needed to achieve substantial time-reflection at optical frequencies.

Here, we study time-reflection and time-refraction of waves incident upon an interface with a dielectric medium experiencing a sudden temporal change. We show that a monochromatic wave incident upon such an interface transforms into pulses. We find that the time-reflection is enhanced near total internal reflection (TIR), especially at the critical angle which acts as an Exceptional Point [38–40]. For incidence above the critical angle, we find that the evanescent wave penetrating into the time-varying medium is transformed into time-refracted and time-reflected propagating (non-evanescent) pulses. The process described here can be applied to analyze various photonic systems involving spatial interfaces with suddenly changing dielectric media, and can enable the first experimental demonstration of a PTC at optical frequencies.

Let us begin with the standard case of a monochromatic plane wave at frequency $\omega$ propagating in a one-dimensional homogenous medium that at $t = 0$ experiences a sudden temporal change in its refractive index from $n_1$ to $n_2$. Maxwell's equations dictate boundary conditions at $t = 0$ such that the electric displacement and magnetic flux density vectors are continuous in time [9]. These conditions give rise to backward and forward propagating waves for $t > 0$. Since the medium is homogeneous, momentum is conserved, hence both of those waves have the same wavevector as the original wave, but at different frequencies $\omega'_{\pm} = \pm\omega\left(\frac{n_1}{n_2}\right)$. The forward wave is the time-refraction and has positive frequency, whereas the backward wave is the time-reflection, which has "negative" frequency. The physical consequences of the negative frequency are that the time-reflected wave propagates backwards with a conjugate phase. For simplicity, we assume that the electric displacement reacts instantaneously to the change in the electric field. Under this assumption, the refraction and reflection coefficients (for amplitudes) for an instantaneous index change (a temporal interface) are $t_{time} = \frac{1}{2}\left(\left(\frac{n_1}{n_2}\right)^2 + \frac{n_1}{n_2}\right)$ and $r_{time} = \frac{1}{2}\left(\left(\frac{n_1}{n_2}\right)^2 - \frac{n_1}{n_2}\right)$ [9,41,44]. Thus, to obtain significant time-reflection, the change in refractive index has to be significant.

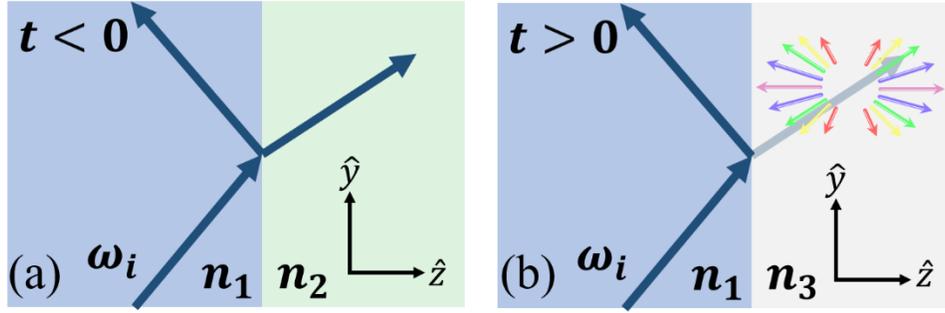

Fig. 1: (a) A monochromatic plane-wave incident upon a spatial interface undergoes refraction and reflection while conserving its original frequency $\omega_i$. (b) A monochromatic plane-wave incident upon a spatial interface with a medium that at $t = 0$ undergoes an abrupt change in its refractive index from $n_2$ to $n_3$. The abrupt index change gives rise to the appearance of a continuous spectrum of waves, where each frequency propagates at a different angle such that higher frequencies are concentrated along the z horizontal axis. Thus, both the time-refracted and the time-reflected waves form spatio-temporal wavepackets which spread during their propagation.

Next, we explore the nature of those phenomena for a monochromatic plane-wave incident upon a spatial interface with a time-varying medium. To do that properly, consider first the simple example of a spatial interface between two media. The interface is in the $xy$ plane, where for $z < 0$ the refractive index is $n_1$ and for $z > 0$ the index is $n_2$. At time $t = 0$ the second medium experiences a sudden temporal change such that its refractive index changes to $n_3$. We denote $\theta_{c,1}$ and $\theta_{c,2}$ as the critical angles for TIR before and after the temporal change, respectively. In this setting, a plane-wave of frequency $\omega_i$, wave-vector $\vec{k}_i = (0, k_y, k_z)$ and amplitude $E_i = 1$ is incident upon this interface at angle $\theta_i$, as sketched in Fig 1. To calculate the evolution of the EM fields for $t > 0$, we carry out the following process. Using the known fields at $t = 0^-$ and the temporal boundary conditions, we find the EM fields at $t = 0^+$, project the fields onto the eigenmodes of the system after the sudden change, evolve each eigenmode in time separately, and reconstruct the total EM fields using superposition. This process holds for every wave system that varies abruptly in time. We apply this process on the example described above for a TE polarized plane-wave, and for simplicity set $n_3 = n_1$. We focus on the region $z > 0$. After the sudden temporal change, the waves are characterized by their wave-vectors and the dispersion $\omega(\vec{k}) =$ $\pm \frac{c}{n_1} |\vec{k}|$. By projecting the fields on the eigenmode basis, we find the amplitudes of the time-reflected waves $E^-(\vec{\kappa})$, which correspond to the negative branch of the dispersion curve, and the amplitudes of the time-refracted waves $E^+(\vec{\kappa})$ which correspond to the positive branch, as described by

$$E^\pm(\kappa_z, \kappa_y) = \frac{t_F}{2}\left(\left(\frac{n_2}{n_1}\right)^2 \pm \frac{\omega(\vec{\kappa})}{\omega_i}\frac{\kappa_y k_y + \kappa_z \beta}{\kappa_y^2 + \kappa_z^2}\right)\left(\frac{1}{i(\beta - \kappa_z)} + \pi\delta(\kappa_z - \beta)\right)\delta(\kappa_y - k_y). \quad (1)$$

We denote $t_F$ as the transmission coefficient and $\beta = \frac{\omega_i}{c}\sqrt{n_2^2 - n_1^2 \sin^2\theta_i}$ as the propagation constant of the original wave in z direction. The critical angle for TIR occurs when the square root is zero, above it $\beta$ is imaginary. The amplitudes of the time-reflected waves are shown in Fig 2(a) for $n_1 = n_3 = 0.8$ and $n_2 = 0.3$. An index change of this magnitude is feasible, as demonstrated recently in epsilon near zero martials [3-7,36]. The symmetric curves shown in Fig 2(a) are explained as follows: for $t < 0$, above $\theta_{c,1}$ the EM field in the region $z > 0$ is an evanescent wave, and as such it is spatially localized very close to the interface. The strong variation in the refractive index at $t = 0$ causes time-reflection, but the wave has no spatial preference, hence the time-reflected waves evolve in the same manner towards both sides of the interface. We see that Eq. (1) displays high sensitivity in the vicinity of $\theta_{c,1}$, due to the transition from a purely real to a purely imaginary value of $\beta$. Prior to the abrupt change in refractive index, the imaginary propagation constant leads to a complex polarization of the magnetic field, resulting in a relative phase of $\frac{\pi}{2}$ between the transverse components of the electric and magnetic fields. As a consequence, the time-average Poynting vector in the z-direction, indicative of photon flux, is zero. However, following the sudden change, a mismatch between the new impedance of the medium and the wave impedance (the ratio between the electric and magnetic fields) is created, altering the relative phase between the fields and leading to a non-zero photon flux in the z-direction. Time-reflection is then induced to reconcile this phase difference and impedance mismatch. The enhancement of the time-reflection near the critical angle for TIR and the sensitivity are similar to other enhancement effects observed at the vicinity of that angle [42,43], because that angle is an Exceptional Point.

Notice that each time-reflected and time-refracted plane-wave has a different frequency according to the dispersion curve, and propagates at a different angle, as sketched in Fig 1(b). The relation between the propagation angle and the frequency can be extracted from momentum conservation in the y direction, which Eq. (1) manifests for the special case $n_3 = n_1$ as the delta function dictates the same $k_y$ for all waves, and is given by

$$\omega(\theta) \sin\theta \, n_3 = \omega_i \sin\theta_i \, n_1. \tag{2}$$

We see from Eq. (2) that, for a given input wave at $\omega_i$ and $\theta_i$, the larger the ratio $\frac{n_1}{n_3}$ - the larger the frequency shift $\omega(\theta) - \omega_i$, and the more the time-refracted and time-reflected spatio-temporal wavepackets spread in space as the relative angles of propagation between waves with adjacent frequencies increase. Notably, above $\theta_{c,1}$, the higher the angle of incidence, the broader the spectrum of plane-waves (Fig 2(a)). Waves with higher $\kappa_z$ have higher frequencies, thus the spectrum of the time-reflection and time refraction is broader. For higher values of $\frac{n_2}{n_1}$, $\beta$ increases, resulting in a wider bandwidth.

The abrupt index change transforms the evanescent wave into a propagating wave, for both the time-refracted and the time-reflected waves. We use Parseval's theorem to estimate the power of the time-reflected and time-refracted waves, displayed in Fig 2(b). The power fraction that is time-reflected is highly enhanced above the critical angle, exceeding 100% of the power of the original wave. For waves incident below $\theta_{c,1}$, the time-reflected power fraction is the same as the reflected power for a plane wave propagating in a homogenous medium whose refractive index is changed abruptly from 0.3 to 0.8 (lower dashed black line in Fig 2(b)), approximately 9.8% of the power of the original wave. Hence, below the critical angle, the time-reflected power from the interface with a medium undergoing a step-change in its refractive index coincides with the simple case of time-reflection of a plane wave in a homogeneous medium, with the reflection coefficient given above. This means that the time-reflected power is enhanced above the critical angle, enhancement that can even reach an order of magnitude compared to the time-reflection below the critical angle.

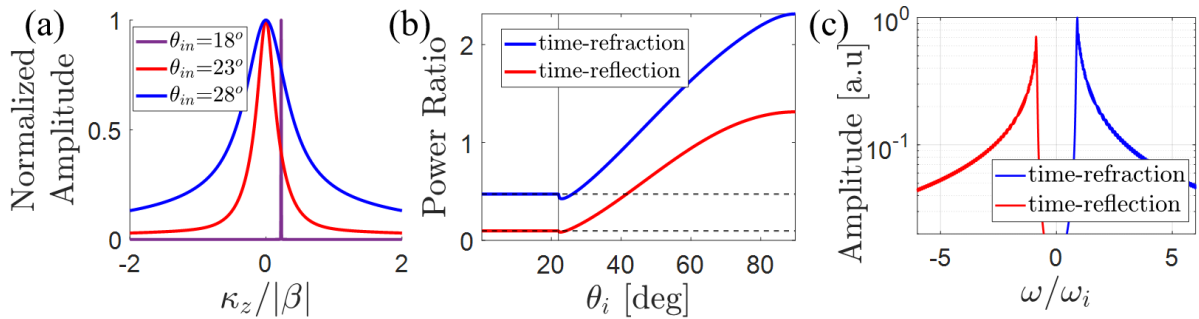

Fig. 2: **Analytic results of the time-refraction and time-reflection induced by a sudden change of a dielectric interface for $n_3 = n_1 = 0.8$ and $n_2 = 0.3$.** The critical angle for TIR is $\theta_{c,1} = 22^o$. (a) Normalized amplitudes of the time-reflected waves vs normalized propagation constant in the $z$ direction, for 3 different incidence angles. Below $\theta_C$ (blue) the Fresnel-refracted wave has a real propagation constant, hence so does the time-reflected wave, and the spectrum is a delta function commensurate with Snell's law. Above $\theta_{c,1}$ the propagation constant of the time-reflected wave is complex, always localized at normal incidence, and has a broader spectrum as the angle is increased above TIR. (b) Ratio between the time-reflected and time-refracted powers to the incident power (defined as the power in the region $z > 0$ at $t < 0$), as a function of the angle of incidence. $\theta_{c,1}$ is marked by a vertical black line. (c) Spectrum of the time-refraction and time-reflection for angle of incident $\theta_i = 50^o$.

Up to this point, we provided analytical analysis of the case in which the optical properties of the system undergo an abrupt change, after which the entire space is homogeneous. We proceed with a numerical study on the time-reflection of a pulse incident on the same dielectric interface, by numerically solving Maxwell's equations for different scenarios. For the simulation, we employ the 2D Finite-Difference Time-Domain (FDTD) method. We create a simulation area, with a dielectric interface at $z = 0$ between $n_1 = 1.5$ at $z < 0$ and $n_2 = 0.3$ for $z > 0$. We launch a TE-polarized pulsed Gaussian beam with a mean frequency $\omega_i = 2 \cdot 10^{15} \frac{rad}{s}$ toward the interface, as illustrated (Fig 3(a)). At time $t = 0$, when the peak of the pulse reaches the interface, we change the refractive index in the region $z > 0$ to $n_3 = 0.8$, such that the critical angle becomes smaller, $\theta_{c,2} < \theta_{c,1} < \theta_i$. We choose the indices of refraction according to recent experiment [36] in which time-refraction was observed. To obtain meaningful information, we choose points in space (A, B, C) through which the incident, the time-reflected and the time-refracted pulses, pass and sample the pulses in time. Using these time samples, we determine the spectrum of each pulse. The simulation results, along with the spectra of the incident pulse, time-reflection and time-refraction, are presented in Fig 3(a) and Fig 3(c) respectively.

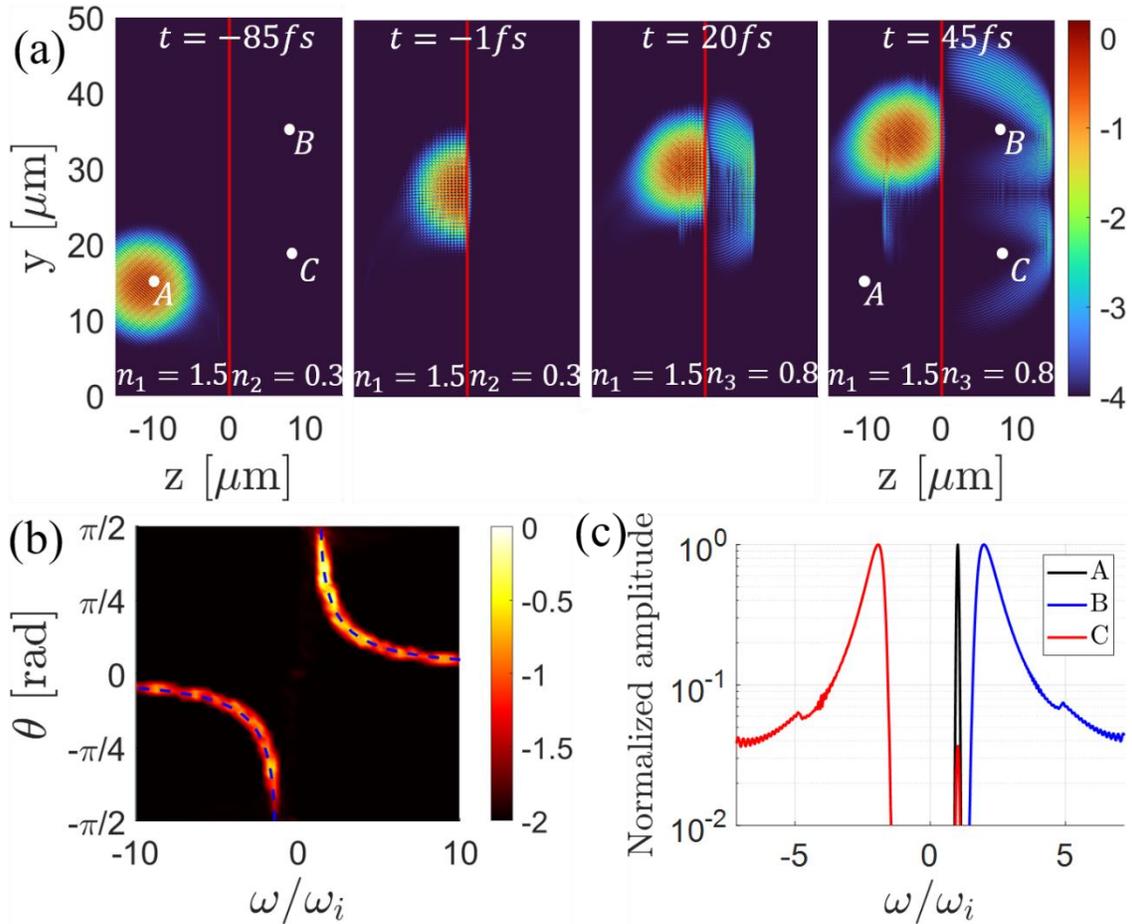

Fig. 3: **Simulations showing the time-refraction and time-reflection of a pulse, induced by a sudden change of a dielectric interface for $n_1 = 1.5, n_2 = 0.3$ and $n_3 = 0.8$.** (a) A pulsed Gaussian beam is launched toward a dielectric interface positioned at $z = 0$ (vertical red line). The EM fields at points A, B and C are being sampled as the simulation evolves in time. The pulse is incident at incidence angle of $\theta_i = 50^o$, which is above $\theta_{c,1}$, hence at $t < 0$ the photon flux in the $z$ direction is zero for $z > 0$. At $t = 0$ the refractive index in the region $z > 0$ is changed to $n_3 = 0.8$, so the angle of incident remains above the new critical angle, $\theta_{c,2}$. (b) Normalized amplitudes (log scale) of waves that composes the electric field in the region $z > 0$ at time $t = 45\,fs$ vs their frequency and angle of propagation, the theoretical curve according to equation (2) is marked by blue dashed line. (c) Normalized spectrum of the incident pulse (black) sampled at the point A in (a), along with the spectra of the time-refracted (blue) and time-reflected (red) pulses sampled at the points B and C in (e). The color bar in (a) represents light intensity (log scale), normalized to the peak intensity of the original pulse.

By examining the result of the simulation, we find 5 pulses after the sudden change in the reflective index, as shown in Fig 3(a). The original pulse is reflected by the temporal change because after the sudden change it stays above the critical angle of TIR. There are two time-refracted pulses which have positive propagation constant in the y direction. One of the time-refracted pulses overlaps with the Fresnel-reflection of the original pulse from the spatial interface, while the other passes through point C. Lastly, we see two time-reflected pulses, one of them passes through point B, while the other propagates toward point A. Before the abrupt change we find zero photon flux in the z direction in the region $z > 0$. After the abrupt temporal change we find that the index variation gives rise to non-zero photon flux in the z direction, carried by the time-reflected and time-refracted pulses). We see that the time-reflection and time-refraction are constructed of waves of different frequencies, each propagates in a different direction, Fig 3(a) at time $t = 45\,fs$. Figure 3(b) shows the wave amplitudes as a function of their propagation angle and frequency along with our theoretical curve following Eq. (2). To analyze the fields recorded at points A, B and C we calculate their spectra, Fig 3(c). The spectra of the time-reflected and time-refracted pulses have similar shapes, differing due to slight asymmetry between point B, C and the center of the pulse, and are broad-banded with respect to the initial pulse. A simple estimation reveals that this broadening of the bandwidth cannot solely be attributed to the bandwidth of the initial pulse. In a homogeneous time-boundary scenario, the new bandwidth would be squeezed by a factor of $\frac{n_2}{n_3} \approx 0.4$, significantly smaller than the observed broadening in Fig 3(c). Thus, this broadening is affected strongly by the spatial interface. To further support our theoretical analysis, we simulate a case of a CW laser beam which is closer in nature to a plane wave. The results are presented in the section B of [44].

Finally, to complement the study, we simulate several chosen cases of the time-varying interface. The scenarios, described in section A of [44] and the figures therein, are similar to that of Fig. 3, with the abruptly-varied refractive indices causing a change in the critical angle of TIR from $\theta_{c,1}$ to $\theta_{c,2}$. In addition to the case presented in Fig 3 where $\theta_{c,2} < \theta_{c,1} < \theta_i$, we examine three additional generic cases. (i) $\theta_{c,1} < \theta_i < \theta_{c,2}$, (ii) $\theta_{c,1} < \theta_{c,2} < \theta_i$, and (iii) $\theta_{c,2} < \theta_i < \theta_{c,1}$, Fig. 4(c). We observe that the abrupt change in the refractive index transforms the incident monochromatic wave propagating (non-evanescent) time-reflected and time-refracted pulses, each carrying energy in the z direction. Generally, we notice that for incidence above the critical angle with a larger ratio $n_2/n_1$, the spectra of the time-reflection and time-refraction pulses are broader, and the higher the ratio $n_1/n_3$ the higher their central frequencies are, as expected from the analysis following Eq. (2). In addition, we examine cases of slower variations of the refractive index, reaching a rate of the single cycle of the EM pulse, section C in [44]. We find that the time-reflection is strongly affected by the variation time of the refractive index, being much stronger for abrupt changes. Nevertheless, the main features of the time-refraction and time-reflection described here are observable for non-instantaneous variation of the refractive index.

To conclude, we studied the reflection and refraction of EM waves from a dielectric interface with a medium undergoing an abrupt change in the refractive index, and focused on the effects near TIR. We found that monochromatic waves are transformed into time-refracted and time-reflected multi-spectral waves that form pulses. Beyond the critical angle for TIR, the evanescent waves are transformed into propagating pulses and the time-reflection is enhanced. The concepts discussed here can be extended to reflection and refraction by a spatio-temporal interface [45, 46], and are expected to display similar results. Our findings raise several intriguing questions. For example, is it possible to design an optical structure that enhances time-reflection and reduces time-refraction? Do spatial evanescent modes exist in a PTC? How does TIR affect the Floquet modes at an interface of PTC and a dielectric medium? Finally, this work helps design the experimental scheme for the observation of time-reflection at optical frequencies.

This work was supported by the US Air Force Office for Scientific Research and by the Breakthrough Program of the Israel Science Foundation.

# References

[1]     F. R. Morgenthaler, *Velocity Modulation of Electromagnetic Waves*, IRE Transactions on Microwave Theory and Techniques **6**, 167 (1958).

[2]     N. Kinsey, C. DeVault, J. Kim, M. Ferrera, V. M. Shalaev, and A. Boltasseva, *Epsilon-near-Zero Al-Doped ZnO for Ultrafast Switching at Telecom Wavelengths*, Optica, OPTICA **2**, 616 (2015).

[3]     L. Caspani et al., *Enhanced Nonlinear Refractive Index in $\varepsilon$-Near-Zero Materials*, Phys. Rev. Lett. **116**, 233901 (2016).

[4]     M. Z. Alam, I. De Leon, and R. W. Boyd, *Large Optical Nonlinearity of Indium Tin Oxide in Its Epsilon-near-Zero Region*, Science **352**, 795 (2016).

[5]     Y. Zhou, M. Z. Alam, M. Karimi, J. Upham, O. Reshef, C. Liu, A. E. Willner, and R. W. Boyd, *Broadband Frequency Translation through Time Refraction in an Epsilon-near-Zero Material*, Nat Commun **11**, 1 (2020).

[6]     O. Reshef, I. De Leon, M. Z. Alam, and R. W. Boyd, *Nonlinear Optical Effects in Epsilon-near-Zero Media*, Nat Rev Mater **4**, 8 (2019).

[7]     V. Bruno, S. Vezzoli, C. DeVault, E. Carnemolla, M. Ferrera, A. Boltasseva, V. M. Shalaev, D. Faccio, and M. Clerici, *Broad Frequency Shift of Parametric Processes in Epsilon-Near-Zero Time-Varying Media*, Applied Sciences **10**, 4 (2020).

[8]     F. Biancalana, A. Amann, A. V. Uskov, and E. P. O'Reilly, *Dynamics of Light Propagation in Spatiotemporal Dielectric Structures*, Phys. Rev. E **75**, 046607 (2007).

[9]     J. T. Mendonça and P. K. Shukla, *Time Refraction and Time Reflection: Two Basic Concepts*, Phys. Scr. **65**, 160 (2002).

[10]    G. Günter et al., *Sub-Cycle Switch-on of Ultrastrong Light–Matter Interaction*, Nature **458**, 7235 (2009).

[11]    C. M. Wilson, G. Johansson, A. Pourkabirian, M. Simoen, J. R. Johansson, T. Duty, F. Nori, and P. Delsing, *Observation of the Dynamical Casimir Effect in a Superconducting Circuit*, Nature **479**, 7373 (2011).

[12]    M. Halbhuber, J. Mornhinweg, V. Zeller, C. Ciuti, D. Bougeard, R. Huber, and C. Lange, *Non-Adiabatic Stripping of a Cavity Field from Electrons in the Deep-Strong Coupling Regime*, Nat. Photonics **14**, 11 (2020).

[13]    B. Apffel, S. Wildeman, A. Eddi, and E. Fort, *Experimental Implementation of Wave Propagation in Disordered Time-Varying Media*, Phys. Rev. Lett. **128**, 094503 (2022).

[14]    M. Lyubarov, Y. Lumer, A. Dikopoltsev, E. Lustig, Y. Sharabi, and M. Segev, *Amplified Emission and Lasing in Photonic Time Crystals*, Science **377**, 425 (2022).

[15]    A. Dikopoltsev, Y. Sharabi, M. Lyubarov, Y. Lumer, S. Tsesses, E. Lustig, I. Kaminer and M. Segev, Light Emission by Free Electrons in Photonic Time-Crystals, Proc. Natl. Acad. Sci. U. S. A., vol. 119, (2022)

[16]    J. R. Zurita-Sánchez, P. Halevi, and J. C. Cervantes-González, *Reflection and Transmission of a Wave Incident on a Slab with a Time-Periodic Dielectric Function $\epsilon(t)$*, Phys. Rev. A **79**, 053821 (2009).

[17]    J. R. Zurita-Sánchez, J. H. Abundis-Patiño, and P. Halevi, *Pulse Propagation through a Slab with Time-Periodic Dielectric Function $\varepsilon(t)$*, Opt. Express, OE **20**, 5586 (2012).

[18]    J. R. Reyes-Ayona and P. Halevi, *Observation of Genuine Wave Vector (k or β) Gap in a Dynamic Transmission Line and Temporal Photonic Crystals*, Applied Physics Letters **107**, 074101 (2015).

[19]    J. Sloan, N. Rivera, J. D. Joannopoulos, and M. Soljačić, *Controlling Two-Photon Emission from Superluminal and Accelerating Index Perturbations*, Nat. Phys. **18**, 1 (2022).

[20]    E. Galiffi, R. Tirole, S. Yin, H. Li, S. Vezzoli, P. A. Huidobro, M. G. Silveirinha, R. Sapienza, A. Alù, and J. B. Pendry, *Photonics of Time-Varying Media*, AP **4**, 014002 (2022).

[21]    S. A. R. Horsley and J. B. Pendry, *Quantum Electrodynamics of Time-Varying Gratings*, Proceedings of the National Academy of Sciences **120**, e2302652120 (2023).

[22]    E. Lustig, Y. Sharabi, and M. Segev, *Topological Aspects of Photonic Time Crystals*, Optica **5**, 1390 (2018).

[23]    V. Pacheco-Peña and N. Engheta, *Temporal Aiming*, Light Sci Appl **9**, 1 (2020).

[24]    Y. Sharabi, E. Lustig, and M. Segev, *Disordered Photonic Time Crystals*, Phys. Rev. Lett. **126**, 163902 (2021).

[25]    V. Pacheco-Peña and N. Engheta, *Temporal Equivalent of the Brewster Angle*, Phys. Rev. B **104**, 214308 (2021).

[26]    E. Lustig, O. Segal, S. Saha, C. Fruhling, V. M. Shalaev, A. Boltasseva, and M. Segev, *Photonic Time-Crystals - Fundamental Concepts*, Opt. Express, **31**, 9165 (2023).

[27]    S. Saha, O. Segal, C. Fruhling, E. Lustig, M. Segev, A. Boltasseva, and V. M. Shalaev, *Photonic Time Crystals: A Materials Perspective*, Opt. Express, **31**, 8267 (2023).

[28]    Y. Sharabi, A. Dikopoltsev, E. Lustig, Y. Lumer, and M. Segev, *Spatiotemporal Photonic Crystals*, Optica, **9**, 585 (2022).

[29]    V. Bacot, M. Labousse, A. Eddi, M. Fink, and E. Fort, *Time Reversal and Holography with Spacetime Transformations*, Nature Phys **12**, 10 (2016).

[30]    Z. Dong, H. Li, T. Wan, Q. Liang, Z. Yang, and B. Yan, *Quantum Time Reflection and Refraction of Ultracold Atoms*, Nat. Photon. (2023).

[31]    O. Y. Long, K. Wang, A. Dutt, and S. Fan, *Time Reflection and Refraction in Synthetic Frequency Dimension*, Phys. Rev. Res. **5**, L012046 (2023).

[32]    H. Moussa, G. Xu, S. Yin, E. Galiffi, Y. Ra'di, and A. Alù, *Observation of Temporal Reflection and Broadband Frequency Translation at Photonic Time Interfaces*, Nat. Phys. **19**, 6 (2023).

[33]    X. Wang, M. S. Mirmoosa, V. S. Asadchy, C. Rockstuhl, S. Fan, and S. A. Tretyakov, *Metasurface-Based Realization of Photonic Time Crystals*, Sci. Adv. **9**, eadg7541 (2023).

[34]    T. R. Jones, A. V. Kildishev, and D. Peroulis, *Time-Reflection of Microwaves by a Fast Optically-Controlled Time-Boundary*, (2023).

[35]    E. Lustig, S. Saha, E. Bordo, C. DeVault, S. N. Chowdhury, Y. Sharabi, A. Boltasseva, O. Cohen, V. M. Shalaev, and M. Segev, *Towards Photonic Time-Crystals: Observation of a Femtosecond Time-Boundary in the Refractive Index*, in *2021 Conference on Lasers and Electro-Optics (CLEO)* (2021), pp. 1–2.

[36]    E. Lustig et al., *Time-Refraction Optics with Single Cycle Modulation*, Nanophotonics **12**, 2221 (2023).

[37]    R. Tirole, S. Vezzoli, E. Galiffi, I. Robertson, D. Maurice, B. Tilmann, S. A. Maier, J. B. Pendry, and R. Sapienza, *Double-Slit Time Diffraction at Optical Frequencies*, Nat. Phys. **19**, 7 (2023).

[38]    S. Klaiman, U. Günther, and N. Moiseyev, *Visualization of Branch Points in PT-Symmetric Waveguides*, Phys. Rev. Lett. **101**, 080402 (2008).

[39]    C. E. Rüter, K. G. Makris, R. El-Ganainy, D. N. Christodoulides, M. Segev, and D. Kip, *Observation of Parity–Time Symmetry in Optics*, Nature Phys **6**, 3 (2010).

[40]    O. Peleg, M. Segev, G. Bartal, D. N. Christodoulides, and N. Moiseyev, *Nonlinear Waves in Subwavelength Waveguide Arrays: Evanescent Bands and the ``Phoenix Soliton''*, Phys. Rev. Lett. **102**, 163902 (2009).

[41]    L. Bar-Hillel, A. Dikopoltsev, Y. Sharabi, E. Lustig, A. Shmuel, and M. Segev, *Time-Reflection beyond the Critical Angle*, in *Conference on Lasers and Electro-Optics (2022), Paper FTh5A.6* (Optica Publishing Group, 2022).

[42]    H. Herzig Sheinfux, I. Kaminer, Y. Plotnik, G. Bartal, and M. Segev, *Subwavelength Multilayer Dielectrics: Ultrasensitive Transmission and Breakdown of Effective-Medium Theory*, Phys. Rev. Lett. **113**, 243901 (2014).

[43]    H. H. Sheinfux, Y. Lumer, G. Ankonina, A. Z. Genack, G. Bartal, and M. Segev, *Observation of Anderson Localization in Disordered Nanophotonic Structures*, Science **356**, 953 (2017).

[44]    See Supplemental Material online.

[45]    S. A. R. Horsley and J. B. Pendry, *Quantum Electrodynamics of Time-Varying Gratings*, Proceedings of the National Academy of Sciences **120**, e2302652120 (2023).

[46]    Z. Li, X. Ma, A. Bahrami, Z.-L. Deck-Léger, and C. Caloz, *Generalized Total Internal Reflection at Dynamic Interfaces*, Phys. Rev. B **107**, 115129 (2023).

Supplementary Information for

# Time-refraction and time-reflection above critical angle for total internal reflection

Lior Bar-Hillel[1+], Alex Dikopoltsev[2,3], Amit Kam[2], Yonatan Sharabi[2], Ohad Segal[1],
Eran Lustig[2] and Mordechai Segev[1,2]

1. *Department of Electrical and Computer Engineering, Technion, Haifa 32000, Israel*
2. *Physics Department, Technion, Haifa 32000, Israel*
3. *Institute of Quantum Electronics, ETH Zurich, 8093 Zurich, Switzerland*

## Section A: Simulations of generic cases

To complement the study presented in the main text, we simulate several select cases of the time-varying interface, Fig 4. The scenarios are similar to that of Fig. 3, with the abruptly-varied refractive indices causing a change in the critical angle of TIR from $\theta_{c,1}$ to $\theta_{c,2}$. In addition to the case presented in Fig 3 where $\theta_{c,2} < \theta_{c,1} < \theta_i$, we now examine three additional generic cases. All those cases are for incidence angle above the critical angle for TIR. When the incidence angle is below the critical angle no evanescent waves are excited, hence the outcome of the abrupt variation of the refractive index is less interesting, amounting to the expansion of the spectrum of the time-refracted and time-reflected waves.

**Case (i)**. $\theta_{c,1} < \theta_i < \theta_{c,2}$. This case is described in Fig 4(a), where after the variation of the refractive index, the incidence angle $\theta_i$ is below the critical angle $\theta_{c,2}$ . As shown in Fig 4(a), the incident pulse experiences ordinary Fresnel-refraction into the medium, and overlaps with the time-refracted pulse. In this case the refractive index increases significantly from $n_2 = 0.3$ to $n_3 = 0.8$, thus, the time-reflection is weak compared to the Fresnel reflection and refraction and compared to the time-refraction. After the sudden change, the refractive indices on both sides of the spatial interface, $n_1 = 0.9$ and $n_3 = 0.8$, are closer in their values. Hence, the bandwidths of the spectra of the time-reflected and time-refracted pulses are narrow, and their mean frequencies are close to the frequency of the incident wave, $\omega_i$ (Fig 4(d) in comparison to other cases (Fig 3(f), 4(e) and 4(f)).

**<u>Case (ii).</u>** $\theta_{c,1} < \theta_{c,2} < \theta_i$. This case is described in Fig 4(b). Here, the incident angle remains above the critical angle of TIR after the sudden temporal change. In this case, $n_3 < n_2$, hence, the time-reflection is more significant than the time-reflection shown in Fig 3(e) and Fig 4(a), and the spectra of the time-reflected and time-refracted pulses, Fig 4(e), are broader than the spectra presented in Fig 3(f). This is a result of a greater difference between $n_1$ and $n_3$.

**<u>Case (iii).</u>** $\theta_{c,2} < \theta_i < \theta_{c,1}$. This case is described in Fig. 4(c). Here, the incident angle is above the critical angle of TIR only after the abrupt temporal change in the refractive index. We see that the time-reflection is significant and has broad spectrum, Figs. 4(c) and 4(f). This is due to large decrease of the refractive index which results in a large difference between $n_3$ and $n_1$. We observe in all the simulations that the time-reflected and time-reflected waves are propagating (non-evanescent) pulses, each carrying energy in the z direction. We notice that, for incidence above the critical angle of TIR with a greater the ratio $n_2/n_1$, the spectra of the time-reflection and time-refraction pulses are broader. Similarly, the higher the ratio $n_1/n_3$ the higher their central frequencies, and the time-reflection and time-refraction pulses spread more widely in space, as anticipated from the analysis following Eq (2).

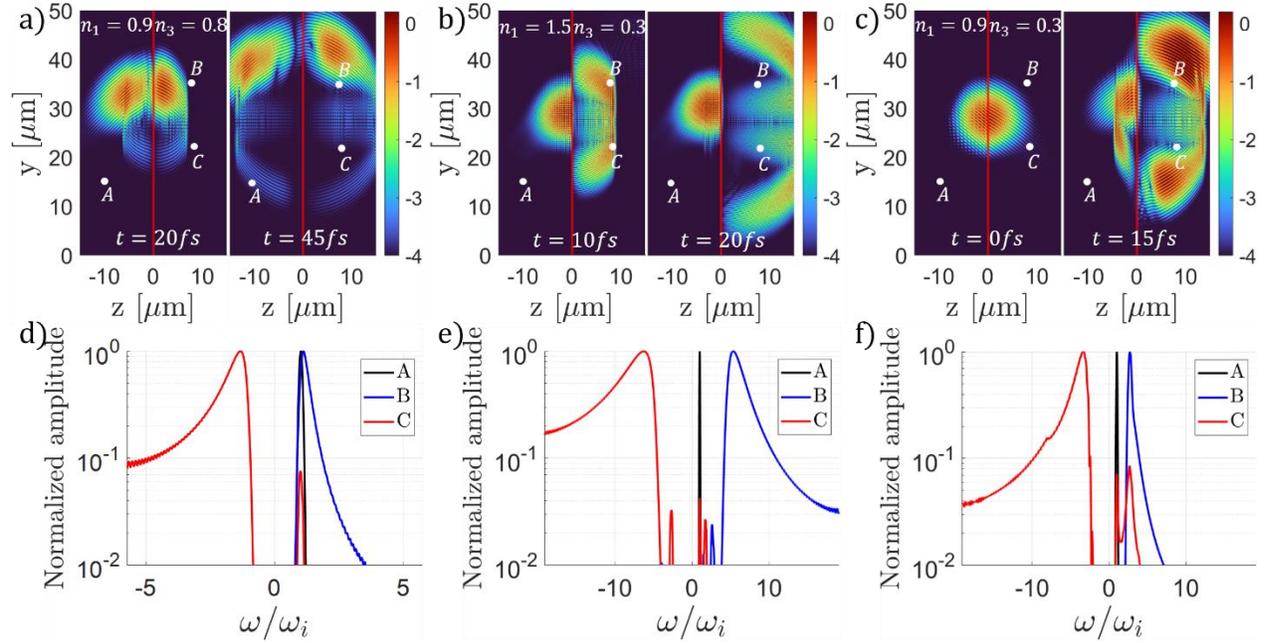

Fig. 4: **FDTD simulations of a pulsed Gaussian beam lunched towards a dielectric interface positioned at $z = 0$ (vertical red line).** The pulsed beam is incident upon the interface at angle $\theta_i = 50^o$. The refractive index in the region $z > 0$ is varied from $n_2$ to $n_3$ when the maximum of the pulse arrives the interface. (a)-(c) Light intensity (log scale) in space after all the pulses propagate away from the interface. The simulated refractive indices ($n_1, n_2, n_3$) are listed above each plot respectively. (d)-(f) Normalized spectrum of the electric field from simulations (a-c) sampled at points A, B and C. The narrow black peak at $\omega = \omega_i$ is the spectrum of the incident pulse measured as it passes through point A. The spectrum of the light in (a) is (d), in (b) is (e) and in (c) is (f).

## Section B: Simulation results for a CW laser beam

To further support our theoretical analysis and observations, we simulate cases of a monochromatic (single frequency) CW laser beam, which is similar in nature to a plane wave. We examine a case similar to that of Fig 3 where $\theta_{c,2} < \theta_{c,1} < \theta_i$, and a case for which $n_3 = n_1$. The results are presented in Fig 5. The small oscillation in the spectra measured are due to the time step (in the simulation) being finite.

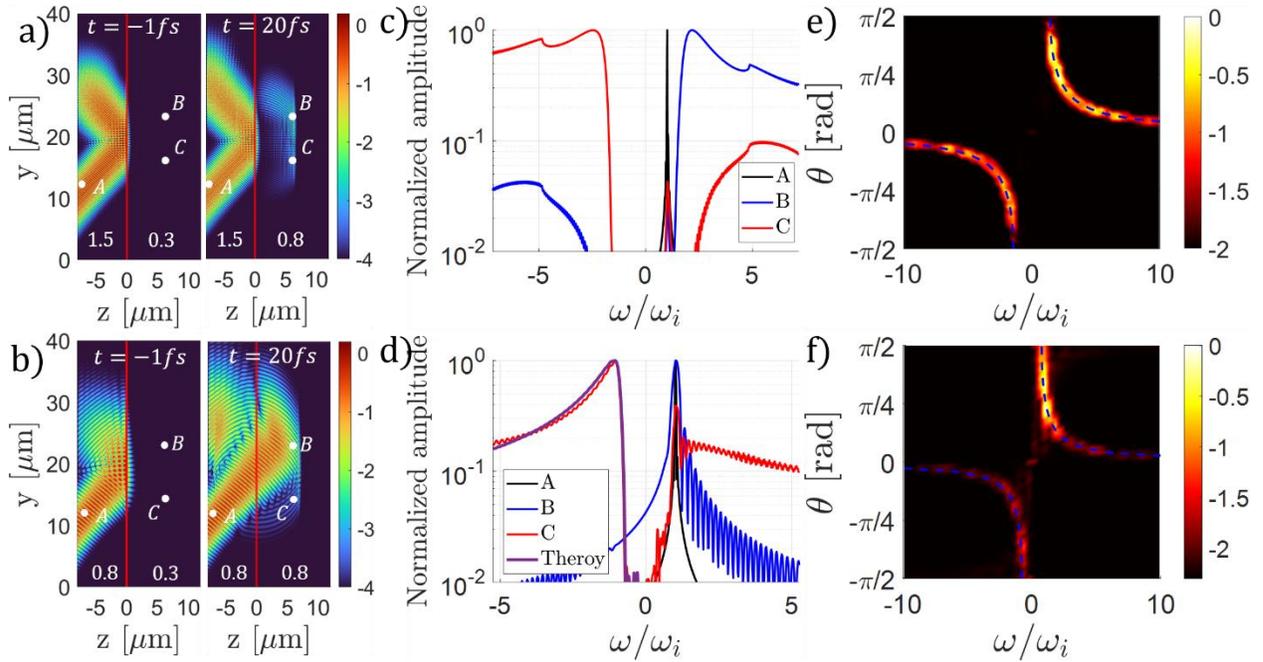

Fig. 5: **FDTD simulations of a single frequency CW laser beam incident upon a dielectric interface positioned at $z = 0$ (vertical red line).** The beam is incident at angle $\theta_i = 50^o$. The refractive index in the region $z > 0$ is varied from $n_2$ to $n_3$ at $t = 0$. The time-refracted (-reflected) waves appear as positive (negative) frequencies in (c-f), respectively.

(a,b) Light intensity (log scale) before the time of the sudden change in the refractive index and after the time-refraction and time-reflections pulses propagate away from the interface. The simulated refractive indices ($n_1, n_2, n_3$) are listed in white at the bottom of each plot.

(c,d) Normalized spectrum of the electric field from simulation (a) and (b), sampled at points A, B and C. The spectrum of the light in (a) is (c) and in (b) is (d). The analytically-calculated spectrum is marked by a purple curve in (d), which coincides with the simulated results for the time-reflection (appearing as negative frequencies in (d)).

(e,f) Normalized amplitudes (log scale) of waves that compose the electric field in (a) and (b) (respectively) in the region $z > 0$ at time $t = 20\,fs$ vs their frequency and angle of propagation. The analytically-calculated curve according to Eq. (2) is marked by blue dashed lines.

We see that the analytically-calculated results (purple curve in Fig 5(d)) align with the simulation results. Likewise, the analytically calculated results (blue dashed lines in Figs. 5(e) and 5(f)) agree well with the simulations in those figures. The positive frequencies observed at point C in Figs. 5(c) and 5(d) arise because the time-reflected pulses at those points did not yet fully separate from the time-refracted pulses. A similar observation applies to points B in Figs. 5(c) and 5(d) for the negative frequencies.

## Section C: Simulations for a dielectric medium whose refractive varies within a single cycle

Recent experiments measured time-refraction within a single cycle modulation time [36]. Thus, it is important to consider feasible cases where the index change is not abrupt but occurring within 1-2 cycles. To do so, we simulate a case for a single frequency CW laser beam incident upon a dielectric interface at $z = 0$, where the index of refraction in the region $z > 0$ varies according to $n(t) = n_2 + (n_3 - n_2) \cdot \frac{1}{2}\left(1 + \tanh\left(\frac{t - \tau/2}{4\tau}\right)\right)$ where $n_2 = 0.3$ and $n_3 = 0.8$. The results are presented in Fig. 6 for different value of $\tau$. We find that the that the time reflection is strongly affected by the non-zero variation time $\tau$. Nevertheless, the main physical features display the same behavior and remain observable even for non-instantaneous changes in the refractive index. As with the abrupt index changes, the evanescent waves transformed into broadband spatio-temporal propagating (non-evanescent) pulses, with a positive photon flux in the $z$ direction. The time-reflected and time-refracted pulses consist of Fresnel waves, each with its own frequency propagating at a different angle, according to Eq. (2).

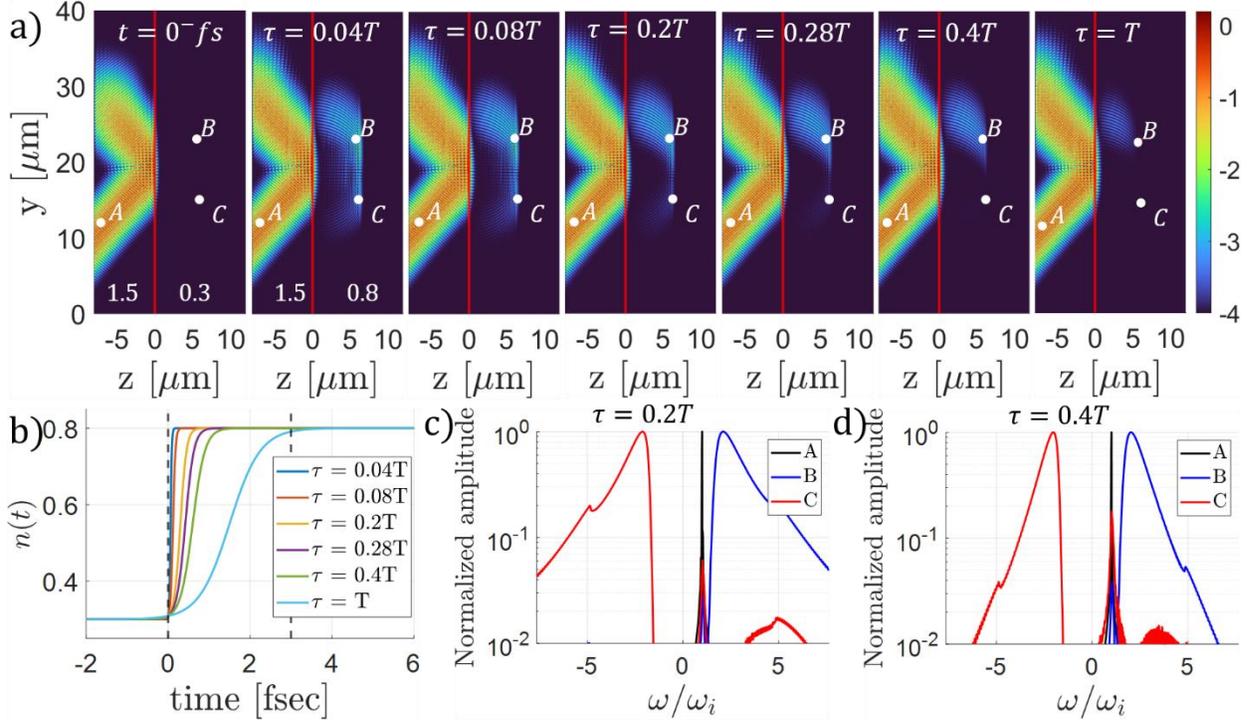

Fig. 6: **FDTD simulations of a single frequency CW laser beam incident upon a time varying dielectric interface positioned at $z = 0$ (vertical red line).** The beam is incident at angle $\theta_i = 50^o$. The refractive index in the region $z > 0$ is varied from $n_2 = 0.3$ to $n_3 = 0.8$ within a time scale $\tau$. (a) Light intensity (log scale) at $t = 20\ fs$. The simulated refractive indices are listed in white at the bottom of the two first plots. Each simulation has different $\tau$. (b) Simulated evolution of the refractive index used in (a). $T$ is the period (one cycle) of the beam, and its duration is marked by the two dashed vertical black line. (c,d) Normalized spectrum of the electric field sampled at points A, B and C in two different simulations with $\tau = 0.2T$ and $\tau = 0.4T$.

From the simulations presented in Fig 6, we observe that the variation time of the refractive index affects mainly the amplitude and the spectrl width of the time-reflection, but not the existence of the main features discussed in the main text and in the examples above. As the index variation takes longer, the amplitude of the time-reflection is smaller its spectrum is narrower. Consequently, the broad bandwidth observed in the time reflection and time refraction, as described in the main text, is a result of the abruptness of the refractive index change.

## Section D: Detailed of the analytic calculation

In this section, we describe the details of the analytic calculations, whose outcome is presented in the main text. We employ the process described in the main text, to find the time reflection and time refraction of EM waves for the special case where after an instantaneous change of the permittivity - the entire space is homogeneous. We denote the Fourier transforms of the electric field and magnetic flux density after the sudden change as $\vec{E}(\vec{k}, t = 0^+)$ and $\vec{B}(\vec{k}, t = 0^+)$. These functions are in momentum space, and they represent the Fourier transforms of the fields using the continuity of $\vec{D}$ and $\vec{B}$. The sum of the time-reflection and time-refraction represents the total field immediately after the abrupt index change, which we implement via mode-matching over all space. We find that

$$\vec{E}^+(\vec{k}) + \vec{E}^-(\vec{k}) = \vec{E}(\vec{k}, t = 0^+),  \tag{3A}$$

$$\frac{\vec{k} \times \vec{E}^+(\vec{k})}{\omega(\vec{k})} - \frac{\vec{k} \times \vec{E}^-(\vec{k})}{\omega(\vec{k})} = \vec{B}(\vec{k}, t = 0^+).  \tag{3B}$$

Multiplying Eq (3B) by $\vec{k} \times$ and using known properties of plane waves (orthogonality of $\vec{k}$ to the fields), the amplitudes of the time-reflection and time-refraction are

$$\vec{E}^\pm(\vec{k}) = \frac{1}{2}\left( \vec{E}(\vec{k}, t = 0^+) \pm \frac{\omega(\vec{k})}{k^2} \vec{k} \times \vec{B}(\vec{k}, t = 0^+) \right).  \tag{4}$$

To implement Eq (4) for the case of a plane wave experiencing a homogeneous instantaneous change of the permittivity from $\varepsilon_1$ to $\varepsilon_2$, we write the Fourier transforms after the abrupt change using the boundary conditions. The Fourier transforms are $\vec{E}(\vec{k}, t = 0^+) = \frac{\varepsilon_1}{\varepsilon_2} E_i \delta(\vec{k} - \vec{k}_i)\hat{x}$ and $\vec{B}(\vec{k}, t = 0^+) = \frac{k_i E_i}{\omega_i} \delta(\vec{k} - \vec{k}_i)\hat{y}$. Using Eq (4) we find the time-reflection coefficient $r = \frac{1}{2}\left( \left(\frac{\varepsilon_1}{\varepsilon_2}\right) - \sqrt{\frac{\varepsilon_1}{\varepsilon_2}} \right)$ and the time-refraction coefficient $t = \frac{1}{2}\left( \left(\frac{\varepsilon_1}{\varepsilon_2}\right) + \sqrt{\frac{\varepsilon_1}{\varepsilon_2}} \right)$. Those coefficients are different from those presented in [9], due to an algebraic error in their derivation. Notably, this derivation is similar for every wave experiencing a homogeneous instantaneous permittivity change.

For the physical scenario of a plane wave incident a time varying dielectric interface presented in the main text, the process is the same. The Fourier transforms of the EM fields in the region $z > 0$ after the sudden change using boundary conditions are $\vec{E}(\vec{\kappa}, t = 0^+) = \frac{\varepsilon_1}{\varepsilon_2} F(\vec{\kappa}) \hat{x}$ and

$$\vec{B}(\vec{\kappa}, t = 0^+) = \frac{F(\vec{\kappa})}{\omega_i} \left( \beta \hat{y} - k_y \hat{z} \right) \quad \text{where} \quad F(\vec{\kappa}) = t_F \left( \frac{1}{i(\beta - \kappa_z)} + \pi \delta(\kappa_z - \beta) \right) \delta(\kappa_y - k_y) \quad .$$

Substituting those expression to Eq (4) gives Eq (1). In this case, the refractive index is discontinuous only in time, whereas for a standard static spatial interface (which yields the known Fresnel coefficients) the refractive index is discontinuous only in the z axis. However, in the scenario described in the main text, the refractive index exhibits a spatiotemporal corner, reflecting discontinuity in both time and space (the t-z plane). The existence of this "corner" in the space-time plane explains the emergence of new temporal frequencies even when a CW (single frequency) is incident, in contradistinction from the spatial-only case of Fresnel coefficients and from the spatially-homogeneous temporally-abrupt case of [9].